\documentclass[12pt]{article}
\usepackage{latexsym}
\usepackage{graphicx}
\usepackage{caption2}
\usepackage{psfrag}
\newcommand{\be}{\begin{equation}}
\newcommand{\ee}{\end{equation}}
\newcommand{\bea}{\begin{eqnarray}}
\newcommand{\eea}{\end{eqnarray}}

\newcommand{\nn}{\nonumber}
\newcommand{\ra}{\rightarrow}

\newcommand{\lesssim}{ {\
\lower-1.2pt\vbox{\hbox{\rlap{$<$}\lower5pt\vbox{\hbox{$\sim$}}}}\
} }
\newcommand{\gtrsim}{ {\
\lower-1.2pt\vbox{\hbox{\rlap{$>$}\lower5pt\vbox{\hbox{$\sim$}}}}\
} }

\input epsf

\begin{document}

\begin{titlepage}

\begin{flushright}
\end{flushright}
\vspace*{1.5cm}
\begin{center}
{\Large \bf An example of resonance saturation at one loop}\\[3.0cm]

{\bf O. Cat\`{a}} and {\bf S. Peris}\\[1cm]

 Grup de F{\'\i}sica Te{\`o}rica and IFAE\\ Universitat
Aut{\`o}noma de Barcelona, 08193 Barcelona, Spain.\\[0.5cm]

\end{center}

\vspace*{1.0cm}

\begin{abstract}

We argue that the large-Nc expansion of QCD can be used to treat
a Lagrangian of resonances in a perturbative way. As an
illustration of this we compute the $L_{10}$ coupling of the
Chiral Lagrangian by integrating out  resonance fields at one
loop. Given a Lagrangian and a renormalization scheme, this is
how in principle one can answer in a concrete and unambiguous
manner questions such as at what scale resonance saturation takes
place.

\end{abstract}

\end{titlepage}

Ever since the early times of Vector Meson
Dominance\cite{Sakurai} there has been constant phenomenological
evidence for the lowest vector and axial vector states to
essentially saturate
 hadronic observables whenever their contribution is allowed by
quantum number conservation. In the context of Chiral Perturbation
Theory \cite{Weinberg,GL} resonance saturation was suggested to
generalize also to the scalar and pseudoscalar
sectors\cite{Swiss1}, and indeed all the $O(p^4)$ $L_i$ couplings
were obtained by means  of integrating out the appropriate
resonance fields\footnote{Ref. \cite{Donoghue} did an analysis
similar in spirit to that of Ref. \cite{Swiss1} where only the
$\rho$ was integrated out.}. However, this integration was
carried out at \emph{tree level}, i.e., the Lagrangian was
effectively treated only as classical.

Specifically Ref. \cite{Swiss1} made the choice to represent
vector and axial-vector particles by antisymmetric tensor fields
and wrote down a Lagrangian with  $SU_3^L\times SU_3^R$-symmetric
interactions of the form\footnote{In this work we shall follow
the same notation as in Ref. \cite{Swiss1}.}
\begin{eqnarray}\label{resonance}
  {\cal L}_{R}&\!\!=\!&-\frac{1}{2}
  {\sum_{R=V,A}}{\langle {\nabla}^\lambda
  R_{\lambda\mu}{\nabla}_\nu
  R^{\nu\mu}- \frac{1}{2} M_R^2 R_{\mu\nu}R^{\mu\nu}}\rangle\,
  +\, \frac{1}{2}\, \langle
  \nabla^\mu S \nabla_\mu S - M_S^2 S^2\rangle  \nn \\
  && +\, \frac{1}{2}\, \langle
  \partial^\mu S_1 \partial_\mu S_1 - M_{S_1}^2 S_1^2\rangle\, +\,
  \frac{F_\pi^2}{4}\, {\langle D_\mu U D^\mu U^\dagger}\rangle \nn \\
  && +\, \frac{F_V}{2\sqrt2}\, {\langle V_{\mu\nu}f_+^{\mu\nu}}\rangle\, +\,
  i\, \frac{G_V}{\sqrt{2}}\, {\langle V_{\mu\nu}u^\mu u^\nu}\rangle\, +\,
  \frac{F_A}{2\sqrt{2}}\, {\langle A_{\mu\nu}f_-^{\mu\nu}}\rangle \nn \\
  && +\, c_d \,\langle S u_{\mu} u^\mu \rangle\, +\, \tilde{c}_d \, S_1\,
  \langle u_{\mu} u^\mu \rangle\, +\,
  L_{10}^\mathcal{^R}\, {\langle U^\dagger F_R^{\mu\nu}UF_{L\mu\nu}}\rangle
  \ ,
\end{eqnarray}
where $V,A,S$ and $S_1$ stand for the octet vector, axial-vector,
scalar  and singlet scalar resonance fields, respectively, and $U$
is the exponential of the Goldstone fields. Other terms appearing
in the Lagrangian of Ref. \cite{Swiss1} will be of no relevance
for the discussion that follows and are not considered in Eq.
(\ref{resonance}).

As is well known the field representation is not unique and, for
instance, in the case of spin-one particles different authors
have chosen different representations to describe them (i.e. an
antisymmetric tensor, Yang-Mills field, Hidden-Symmetry field,
etc...\cite{Yamawaki, Meissner}). As a consequence of this, it
was seen that ambiguities in physical observables may occur. In
Ref. \cite{Swiss2} these ambiguities were resolved by imposing
short-distance matching onto the QCD Operator Product Expansion
of certain Green's functions. As a matter of fact, it was shown
later on in Ref. \cite{BijnensPallante} that all the above
choices in the representation were actually field redefinitions
of the particular Lagrangian of Eq. (\ref{resonance}).

Let us take the case of $L_{10}$ as an example. Integrating the
vector and axial-vector fields in the Lagrangian
(\ref{resonance}) at tree level leads to the low-energy chiral
Lagrangian of Eq. (\ref{chiral}) (see below) with equations
relating couplings below and above threshold, such as
\begin{equation}\label{L10}
  L_{10}(\mu)=\frac{F_A^2}{4M_A^2}-\frac{F_V^2}{4M_V^2}+ L_{10}^\mathcal{R}(\mu)\
  .
\end{equation}
Here $L_{10}(\mu)$ stands for the $O(p^4)$ coupling in the
low-energy Lagrangian \emph{after} the V and A resonance fields
have been integrated out, i.e.
\begin{equation}\label{chiral}
  {\cal L}_{\chi\rm{PT}}=\frac{F_\pi^2}{4}{\langle D_\mu U D^\mu
    U^\dagger}\rangle\,\,+\,\,L_{10}\ {\langle U^\dagger
    F_R^{\mu\nu}UF_{L\mu\nu}}\rangle + \mathcal{O}(L_i,i=1,\ldots,9) \ ,
\end{equation}
whereas $ L_{10}^\mathcal{R}(\mu)$ is the akin coupling, but at
the level of the resonance Lagrangian (\ref{resonance}). The
other couplings $L_{1-9}$ complete the list at
$\mathcal{O}(p^4)$\cite{GL}. The statement of resonance saturation
is then tantamount to the equation
\begin{equation}\label{saturation}
  L_{10}^\mathcal{R}(\mu)=0 \ ,
\end{equation}
and expresses the fact that the whole low-energy coupling $L_{10}$
is directly ``produced" in the process of integrating the
resonance field.

The result  of Eq. (\ref{saturation}) is actually field
representation dependent and is true only in the antisymmetric
tensor formulation, i.e. for the Lagrangian in Eq.
(\ref{resonance}). Other formulations (i.e. Yang-Mills,
Hidden-Symmetry,etc...) may have non-zero values of
$L_{10}^\mathcal{R}(\mu)$ to balance the different contribution
from the direct integration of the resonance fields to finally
produce the same value for $L_{10}(\mu)$, as it is produced by the
field redefinition connecting the different
formulations\cite{BijnensPallante}. Alternatively, this may also
be seen as a consequence of certain matching conditions to QCD at
short distances\cite{Swiss2}. Therefore, although at tree level
$L_{10}(\mu)$ is always given by the same combination of
resonance parameters regardless of the formulation, it is in the
antisymmetric tensor representation that it originates solely
from the interactions of the resonance fields in the Effective
Lagrangian, with the matching to QCD at short distances appearing
as automatic.

Since the left hand side of Eq. (\ref{saturation}) in general
obeys a nontrivial Renormalization Group equation, i.e. it is
$\mu$ dependent, while the right hand side is a constant, this
equation has to be supplied with the prescription of some value
for $\mu$ at which it is supposed to be valid, which we shall
call $\mu^*$\footnote{Unless $L^{\mathcal{R}}_{10}(\mu)=0$
identically $\forall \mu$, of course. We shall comment on this
possibility at the end.}. Notice that if it happens that
$L^{\mathcal{R}}_{i}(\mu^{*})=0$ ($i=1,...,10$), for a certain
$\mu^{*}$ of the order of a resonance mass, then one can use this
as a boundary condition to \emph{predict} all the low-energy
couplings $L_{i}(\mu)$ of the chiral Lagrangian at scales
$\mu\leq \mu^{*}$. The scale $\mu^{*}$ can then be given the
meaning of a threshold between the low-energy chiral Lagrangian
and the resonance Lagrangian that would take over at higher
energies\footnote{This is somewhat similar to the Grand
Unification program, only that at energies which are 15 orders of
magnitude below!}.

In Ref. \cite{Swiss1,Swiss2} it was argued that the natural
choice is $\mu^{*}=M_V$; and the coupling
$L_{10}^\mathcal{R}(M_V)$  was omitted from the resonance
Lagrangian (\ref{resonance}) in accord with Eq.
(\ref{saturation}). With this prescription for $\mu^{*}$ Eq.
(\ref{L10}) leads to a prediction for $L_{10}(M_V)$ in terms of
known resonance masses and decay constants. Similar results were
also obtained for all the rest of the $L_i, i=1,\dots,9$ with
remarkable overall agreement with the experimental
determinations\cite{Swiss1,Amorosetal}.

However the former agreement, although clearly important, is
necessarily only of a qualitative nature. No attempt is made at
defining the underlying QCD approximation that is being used and,
as a consequence, it is not clear how to systematically improve
it. For instance the prescription $\mu=M_V$ to effect resonance
saturation may indeed be natural but only as long as one is
prepared not to distinguish between the two scales
$M_V=0.77\textrm{GeV}$ and  $M_A=1.25\textrm{GeV}$, both of which
in turn must be identified with something like $\Lambda_\chi\sim
1\textrm{GeV}\gg M_{K,\pi}$. At some level of accuracy, however,
one may eventually want to distinguish between $M_V$ and $M_A$;
after all $M_A-M_V$ is actually larger than $M_V-M_K$, for
instance. Furthermore, it is not clear from just a tree-level
integration whether the Lagrangian (\ref{resonance}) actually
saturates the $L_i$'s at $\mu=M_V$, since the scale $\mu$ first
appears at one loop.

In Ref. \cite{PPdeR} it was realized that the above scheme of
resonance saturation can be best understood as  an approximation
to large-$N_c$ QCD\cite{Hooft}, which was called Lowest Meson
Dominance. This is the approximation in which, out of the (in
principle) infinite set of resonances, only the lowest one is
kept in each channel. We remark that this approximation can be
improved upon since, in principle, more resonances may be added
whose couplings and masses can be fixed by matching to higher
terms in the Operator Product Expansion at short distances.
Adopting the large-$N_c$ expansion right from the start
justifies, for instance, the \emph{tree-level} integration of the
resonance fields employed in Ref. \cite{Swiss1,Swiss2} since this
is precisely the leading contribution at large
$N_c$\footnote{Furthermore, assuming that confinement takes place
at large $N_c$, the $1/N_c$ expansion supplies a framework in
which quark and meson degrees of freedom match and no problems of
double counting arise. See the second paper in Ref.
\cite{Hooft}.}. This also tells you that it makes no sense to be
more precise on the value of the scale $\mu$ at which one is
doing the matching unless one goes to the next order in the
large-$N_c$ expansion, as the difference between two scales $\mu$
and $\mu'$ necessarily yields a contribution of subleading order
in $1/N_c$. Consequently, Eq. (\ref{L10}) is a statement at
leading order in the $1/N_c$ expansion in which the $\mu$
dependence of both sides remains, strictly speaking, ill-defined
until next-to-leading (i.e. quantum) effects are
computed\footnote{The situation is somewhat similar to QED:
$\alpha$ only runs with scale after considering quantum effects.}.
This new point of view of resonance saturation as an
approximation to large-$N_c$ QCD is now being studied and
successfully applied to many different problems in hadron
physics\cite{theworks}.

In this letter we shall adopt large-$N_c$ as our underlying
expansion and (\ref{resonance}) as our resonance Lagrangian. We
merely wish to illustrate the point that, as a consequence of the
large-$N_c$ expansion, it \emph{makes} sense to compute quantum
corrections with a resonance Lagrangian and ask, for instance, the
question of at which scale $\mu$ resonance saturation takes
place, if it does at all. Specifically we shall consider $1/N_c$
quantum effects that give rise to a nontrivial $\mu$ dependence in
Eq. (\ref{L10}).

In order to make this explicit we shall take the Lagrangian
(\ref{resonance}) as our starting point\footnote{The advantage of
having a Lagrangian is that, in principle, one can go and compute
quantum corrections with it!}. This we do although this
Lagrangian is probably too simple to  satisfy the short-distance
constraints of QCD at next-to-leading order in the large-$N_c$
expansion, even in the particular case of the $\Pi_{LR}$ function
which will be the relevant one here. Therefore, in this sense,
our analysis cannot be considered fully realistic for QCD. Notice
that Ref. \cite{Swiss2} showed the good matching of this
Lagrangian to QCD only at leading order in $1/N_c$ and, even
then, only for certain Green's functions. Further interesting
studies can be found in \cite{Moussallam} and, in particular, in
\cite{Knecht-Nyffeler}. It is obvious that determining the
resonance Lagrangian that satisfies the short distance
constraints at the next-to-leading order in $1/N_c$, even only in
all the Green's functions studied up to now, is an extremely
arduous task. Therefore, here we will have to content ourselves
with a much more modest goal.

In this letter we shall restrict ourselves to the particular case
of the $L_{10}$ coupling. This we do because this coupling is
defined in terms of a two-point Green's function in QCD, which
makes life simpler. At the same time both vector and axial-vector
particles affect $L_{10}$, which makes it a sensitive probe for
whether $M_V$ or $M_A$ (or neither one) should be the relevant
scale driving the statement of resonance saturation, Eq.
(\ref{saturation}). In other words, we want to find out if the
Lagrangian of Eq. (\ref{resonance}) is at least capable of
reproducing the right value for $L_{10}$ at some scale $\mu^{*}$,
once quantum corrections are taken into account and whether this
scale $\mu^{*}$ indeed coincides with $M_V$ or not. This will
entail a calculation with the Lagrangian of Eq. (\ref{resonance})
and resonances running around in loops. It is then that the
large-$N_c$ counting becomes important. Resonances are not
amenable to a chiral counting like Goldstone bosons are and, were
it not for the large-$N_c$ expansion, there would be no obvious
small parameter with which to do perturbation theory \footnote{In
certain special circumstances one \emph{can} set up a coherent
framework in which resonance loops make sense through a chiral
counting\cite{Manohar}. In general,though, this is not
possible.}. This is the main advantage of the large-$N_c$
expansion for the purposes of this work: QCD in the limit
$N_c\rightarrow\infty$ is a theory of free, noninteracting
hadrons and, consequently, interactions among them are modulated
by increasing inverse powers of $N_c$. In other words,
\emph{there is} a ``small'' coupling governing hadron
interactions (no matter at which energy) and, with it, a sense in
which loops are smaller than the tree level.

One of the consequences of using the large-$N_c$ expansion is
that now we have to enlarge the flavor symmetry in the Lagrangian
of Eq. (\ref{resonance}) from $SU_3^L\times SU_3^R$ to
$U_3^L\times U_3^R$ \cite{largeN} to incorporate the
$\eta_1$\footnote{Since the $\eta_1$ starts playing  a role in
our discussion of $L_{10}$ at $ \mathcal{O}(N_c^0)$ we may
consider it as truly massless. We shall see at the end that our
result depends very little on this, however.}. This can easily be
done by means of the replacement
\begin{equation}\label{grandN}
  U\longrightarrow U e^{-i \frac{\sqrt{2}}{\sqrt{3}}
  \frac{\eta_{1}}{F_\pi}} \ .
\end{equation}

To begin, let us define the $\Pi_{LR}$ function ($Q^2\equiv
-q^2\ge 0$ for $q^2$ space--like) as
\begin{equation}\label{Pilr}
\Pi_{LR}^{\mu\nu}(q)\delta_{ab}=
 2i\int d^4 x\,e^{iq\cdot x}\langle 0\vert
\mbox{\rm T}\left(L^{\mu}_a(x)R^{\nu}_b(0)^{\dagger} \right)\vert
0\rangle\,,
\end{equation}
with color-singlet currents
\begin{equation}\label{currents}
  R_a^{\mu}\left(L_a^{\mu}\right)=
\bar{q}(x)\gamma^{\mu}\ \frac{\lambda_a}{\sqrt{2}}\
\frac{(1\pm\gamma_5)}{2}\ q(x)\, ,
\end{equation}
where $q=u,d,s$ and $\lambda_a$ are Gell-Mann matrices in flavor
space.

 In the chiral limit, $m_{u,d,s}\ra 0$\, , this correlation
function has only a transverse component,
\begin{equation}\label{Ppilr}
  \Pi_{LR}^{\mu\nu}(Q^2)=(q^{\mu}q^{\nu}-g^{\mu\nu}q^2)\Pi_{LR}(Q^2)\,.
\end{equation}

\begin{figure}
\renewcommand{\captionfont}{\small \it}
\renewcommand{\captionlabelfont}{\small \it}
\centering
\includegraphics[width=1.5in]{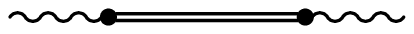}
\hspace{0.5in}
\includegraphics[width=1.5in]{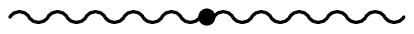}
\caption{{Tree-level contribution to $\Pi(q^2)_{LR}$ from the
Lagrangian in Eq.(\ref{resonance}). The double line stands for
the $V$ and $A$ propagators; the dot for the insertion of
$L_{10}^{\mathcal{R}}$.}}\label{fig:treeresonance}
\end{figure}
\begin{figure}
\renewcommand{\captionfont}{\small \it}
\renewcommand{\captionlabelfont}{\small \it}
\centering
\includegraphics[width=1.5in]{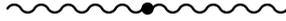}
\caption{Tree-level contribution from the Lagrangian in
Eq.(\ref{chiral}), but now the dot stands for the insertion of
$L_{10}$.}\label{fig:treeL}
\end{figure}

At low energy Green's functions in general, and $\Pi_{LR}(Q^2)$ in
particular, should be equal in the two theories with Lagrangians
(\ref{resonance}) and (\ref{chiral}). For the case of $L_{10}$
that we are here concerned with, it is immediately seen that
 the result in Eq. (\ref{L10}) is the matching condition that results (at tree level)
 from demanding that the ``slope'' in $Q^2$, i.e. the combination
\begin{equation}\label{derivative}
  -\frac{1}{4}\frac{d}{dQ^2}\left\{Q^2
  \Pi_{LR}(Q^2)\right\}_{Q^2=0}\ ,
\end{equation}
be equal when computed both with the Lagrangian in Eq.
(\ref{resonance}) and with that in Eq. (\ref{chiral}). This is
just given by the diagrams in Figs. \ref{fig:treeresonance} and
\ref{fig:treeL} .

\begin{figure}
\renewcommand{\captionfont}{\small \it}
\renewcommand{\captionlabelfont}{\small \it}
\centering
\includegraphics[width=1.5in]{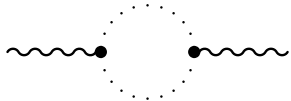}
\caption{One-loop contribution to $\Pi_{LR}$ from the Lagrangian
in Eq.(\ref{chiral}). The dotted line stands for the Goldstone
bosons included in the matrix $U$. }\label{fig:oneloopchiral}

\end{figure}

We now move to the contribution at one loop. Firstly, let us
consider the contribution to $\Pi_{LR}$ stemming from the
Lagrangian in Eq. (\ref{chiral}). The result is given by the
diagram depicted in Fig. \ref{fig:oneloopchiral} plus again the
direct contribution from the coupling $L_{10}$ in Fig.
\ref{fig:treeL}. As a renormalization scheme we shall use
throughout the particular $d$-dimensional
$\overline{\textrm{MS}}$ variant used in \cite{GL} in which, e.g.,
$L_{10}$ renormalizes according to
\begin{equation}\label{MS}
  L^\mathrm{bare}_{10}=L_{10}(\mu)- \frac{1}{4}
  \frac{\mu^{d-4}}{(4\pi)^2}\left\{\frac{1}{d-4}-\frac{1}{2}\left(\log
  4\pi+\Gamma'(1)+1\right)\right\} \ .
\end{equation}
Then one obtains the well-known result
\begin{equation}\label{runningL10}
  \Pi_{LR}(Q^2)= 4 L_{10}(\mu) - \frac{1}{32\pi^2}
  \left(\frac{5}{3}-\log{\frac{Q^2}{\mu^2}}\right) \ ,
\end{equation}
where $L_{10}(\mu)$ is the renormalized coupling in
$\overline{\textrm{MS}}$. Since $\Pi_{LR}$ is $\mu$ independent
this equation implies the usual renormalization group equation for
$L_{10}(\mu)$.

\begin{figure}
\renewcommand{\captionfont}{\small \it}
\renewcommand{\captionlabelfont}{\small \it}
\centering
\includegraphics[width=1.5in]{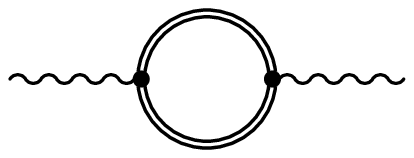}
\includegraphics[width=1.5in]{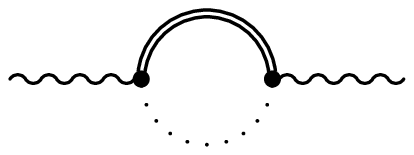}
\includegraphics[width=1.5in]{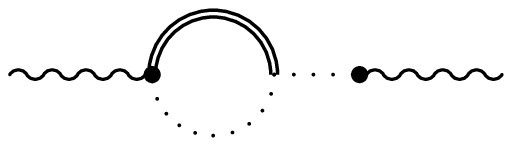}
\caption{One-loop contribution to $\Pi_{LR}$ from the Lagrangian
of resonances in Eq.(\ref{resonance}). Double lines stand for
resonance propagators. Dotted lines stand for Goldstone
propagators. }\label{fig:resonanceloop}
\end{figure}

The diagrams giving the resonance contribution  to $\Pi_{LR}$ at
one loop are depicted in Fig. \ref{fig:resonanceloop}. Adding all
the one-loop contributions in Fig. \ref{fig:resonanceloop} to the
tree-level of Fig. \ref{fig:treeresonance} and to the one-loop of
Goldstones in Fig. \ref{fig:oneloopchiral} one gets the total
contribution to $\Pi_{LR}$ from the Lagrangian (\ref{resonance}).
Equating this expression to that in Eq. (\ref{runningL10}) one
verifies that the Goldstone loop cancels out in the matching (as
it should) and one finally obtains for $L_{10}$

\begin{eqnarray}\label{result}
4\ L_{10}(\mu)&=&\frac{F_A^2}{M_A^2}-\frac{F_V^2}{M_V^2}\nonumber\\
&&-\frac{3}{2}\frac{F_A^2}{f_\pi^2}\frac{1}{(4\pi)^2}\left(\frac{1}{2}-
\log\frac{M_A^2}{\mu^2}\right)+
\frac{3}{2}\frac{F_V^2}{f_\pi^2}\frac{1}{(4\pi)^2}\left(\frac{1}{2}-
\log\frac{M_V^2}{\mu^2}\right)\nonumber\\
&&-\frac{5}{(4\pi)^2}\frac{G_V^2}{f_\pi^2}
\left(-\frac{17}{30}-\log\frac{M_V^2}{\mu^2}\right)\nonumber\\
&&+\frac{3}{2}\frac{1}{(4\pi)^2}\left(-\frac{1}{3}-\log\frac{M_A^2}{\mu^2}\right)
+\frac{3}{2}\frac{1}{(4\pi)^2}\left(-\frac{1}{3}-\log\frac{M_V^2}{\mu^2}\right)
\nonumber\\
&&
-\frac{4}{3}\left(\frac{{\tilde{c}}_d}{f_{\pi}}\right)^2\frac{1}{(4\pi)^2}
\left(\frac{1}{6}+\log\frac{M_{S_1}^2}{\mu^2}\right)-\frac{10}{9}
\left(\frac{c_d}{f_{\pi}}\right)^2\frac{1}{(4\pi)^2}
\left(\frac{1}{6}+\log\frac{M_S^2}{\mu^2}\right)\nonumber\\
&&+\frac{1}{2}\frac{1}{(4\pi)^2}
\left(1+\log\frac{M_S^2}{\mu^2}\right)-\frac{4}{9}\left(\frac{c_d}{f_{\pi}}\right)^2
\frac{1}{(4\pi)^2}\Bigg[\frac{1}{6}+\log\frac{M_S^2}{\mu^2}\nonumber\\
&& + 2B + 2B^2 - (2 B^3 + 3 B^2) \log\frac{M_S^2}{M_{\eta_{1}}^2}\Bigg]\nonumber\\
&&+ 4\ L_{10}^{\mathcal{R}}(\mu) \ ,
\end{eqnarray}
where we have defined $B=M_{\eta_{1}}^2/(M_S^2-M_{\eta_{1}}^2)$.
As to the large-$N_c$ counting, we shall consider
$L_{10}^{\mathcal{R}}(\mu)$ of $\mathcal{O}(1)$\footnote{We
remark that $L_{10}^{\mathcal{R}}(\mu)$ may have contributions of
$\mathcal{O}(N_c)$ stemming from the integration of resonances
(with a mass $M_R$, say) which are even heavier than those
explicitly considered in the Lagrangian (\ref{resonance}).
However these contributions are down by $1/M_R^2$ and we
disregard them here. Whether this is a good approximation or not
will depend on the details of the Lagrangian giving rise to
$L_{10}^{\mathcal{R}}(\mu)$ in Eq. (\ref{resonance}).}. Since
$F_{V,A}^2, G_V^2, c_d^2$ and $\tilde{c}_d^2 $ are
$\mathcal{O}(N_c)$, while the resonance masses are $
\mathcal{O}(1)$, the tree-level contribution above is
$\mathcal{O}(N_c)$ while the one-loop one is $\mathcal{O}(1)$, as
it should.
\begin{figure}
\renewcommand{\captionfont}{\small \it}
\renewcommand{\captionlabelfont}{\small \it}
\centering
\psfrag{m}{$\mu$}
\includegraphics[width=4.5in]{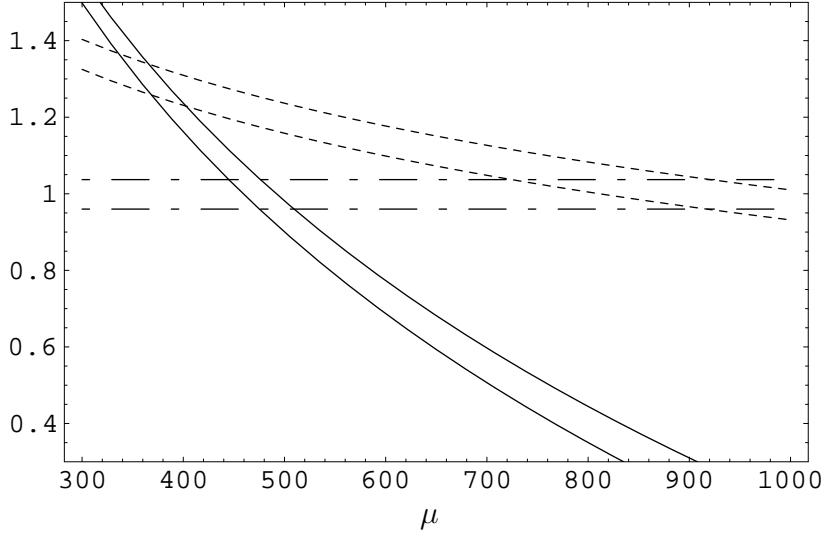}
\caption{This figure shows the curve obtained for $L_{10}(\mu)$
in Eq. (\ref{result}) as a function of $\mu$ (in MeV) under the
condition $L^\mathcal{R}_{10}(\mu)=0$ (solid curve). The dashed
curve is the running of $L_{10}(\mu)$ according to Eq.
(\ref{experiment}). For comparison we also show the tree level
contribution to $L_{10}(\mu)$ (first line in Eq. (\ref{result}))
as dot-dashed lines. All three curves have been normalized to the
central value of the tree level resonance contribution.
}\label{fig:plot}
\end{figure}

In the work of Refs. \cite{Swiss2,PPdeR} it was found that the
tree-level matching (i.e. leading at large $N_c$) of the
Lagrangian (\ref{resonance}) to the short-distance behavior of
certain Green's functions in QCD led to the constraints
$F_A=f_\pi$, $F_V=\sqrt{2} f_\pi$, $G_V=f_\pi/\sqrt{2}$,
$M_A=\sqrt{2} M_V$ and $ M_V= 4\pi f_\pi(\sqrt{6}/5)^{1/2}$.
Consequently these are constraints among the parameters in the
Lagrangian (\ref{resonance}). These are not in principle the same
thing as the physical mass (defined as the pole in the
propagator) and physical decay constant (defined,e.g., through
the width), but it is precisely the parameters in the Lagrangian
and not the physical ones what appears in Eq. (\ref{result}).
This is why the first term in Eq. (\ref{result}), namely
\begin{equation}\label{prediction}
  \frac{F_A^2}{M_A^2}-\frac{F_V^2}{M_V^2} \ ,
\end{equation}
is in fact \emph{predicted} to be
\begin{equation}\label{prediction2}
 -\frac{15}{32 \pi^2\sqrt{6}}\
\end{equation}
in this equation. In passing one also sees that the $\log M_A$
dependence cancels out in Eq. (\ref{result}). It is intriguing to
entertain the idea that the above relations between masses and
decay constants could be a consequence of a higher symmetry of
the planar graphs of QCD.

Looking at (\ref{result}), one sees clearly that knowledge of
$L^{\mathcal{R}}_{10}(\mu)$ immediately translates into a
prediction for $L_{10}(\mu)$. Using the above constraints and
$f_{\pi}=87\pm 3.5$ MeV (chiral limit) together with the
phenomenological values $c_d^2 = \tilde{c}_d^2/3= 1024$ MeV$^2$
and $M_S=M_{S_1}=983$ MeV\cite{Swiss1}\footnote{The final result
is quite insensitive to the precise values of the parameters in
the spin-zero sector. In fact one can change the $c_d$ and
$\tilde{c}_d$ couplings and the scalar masses by a factor of two,
and the $\eta_1$ mass between zero and $980$ MeV without any
dramatic change in the result.}, one can now take Eq.
(\ref{result}) and compare it to the expression for the running
of $L_{10}$\cite{GL}:
\begin{equation}\label{experiment}
  L_{10}(\mu)=L_{10}(M_\rho) - \frac{1}{64 \pi^2}
  \log\frac{M_\rho}{\mu}\ ,
\end{equation}
where $L_{10}(M_\rho)=(-5.13 \pm 0.19)\times 10^{-3}$ \cite{DHGS}.
This comparison is made in Fig. \ref{fig:plot}. In this figure
one can see how the Lagrangian (\ref{resonance}) is actually able
to produce the right experimental value for $L_{10}(\mu)$, but at
a value for $\mu$ which is much lower than what was expected in
Ref. \cite{Swiss1}. This happens at $\mu^{*} \sim 380$ MeV where,
as it turns out, the condition of resonance saturation is
fulfilled, namely $L^{\mathcal{R}}_{10}(\mu^{*})=0$. Notice that,
as Fig. \ref{fig:plot} shows, at the scale $\mu^*$ the one-loop
radiative corrections are $\sim 30\%$ of the tree level, so one is
reasonably within the perturbative regime expected for the
$1/N_c$ expansion. In fact, at $\mu\sim 490$ MeV the one-loop
contribution vanishes altogether. On the other hand, at higher
scales the one-loop corrections quickly grow and one finds, e.g.,
a $\sim 60\%$ reduction relative to the tree level at $\mu\sim
800\ \mathrm{MeV}\sim M_V$; with even larger corrections the
higher the scale $\mu$. In other words, at this scale
$L^{\mathcal{R}}_{10}$ must be $\sim 60\%$ of the tree level and
clearly different from zero for Eq. (\ref{result}) to be
satisfied. Therefore resonance saturation for the Lagrangian
(\ref{resonance}) with the renormalization scheme (\ref{MS}) does
not take place at the large values of $\mu$, namely $\mu \approx
\Lambda_{\chi}\sim 1$GeV, where one would like the resonance
Lagrangian (\ref{resonance}) to take over from the low-energy
chiral Lagrangian (\ref{chiral}).

Perhaps some discussion on the meaning of
$L^{\mathcal{R}}_{10}(\mu^*)=0$ is now in order. As a matter of
fact, only from the knowledge of the value of $\mu^*$ which
satisfies this condition one does not learn much. For one thing
$L^{\mathcal{R}}_{10}(\mu)=0$ is a renormalization scheme
dependent condition on $\mu^*$ (in our case the scheme was given
in Eq. (\ref{MS})) and therefore, strictly speaking, $\mu^*$ can
be shifted by a change in the renormalization scheme
\footnote{This is not strange, matching conditions are also scheme
dependent in the integration of a heavy quark in the running of
$\alpha_{s}(\mu)$, for instance\cite{Match}.}. Even with this
caveat in mind, the low value of $\mu^*$ obtained in Eq.
(\ref{result}) makes one suspect the Lagrangian
(\ref{resonance}), if only because one already knows that
(\ref{resonance}) has several drawbacks, like e.g. the wrong
short-distance behavior of certain Green's functions due to the
lack of a $\pi\rho\mathrm{a_{1}}$ coupling; just to mention one
of them \cite{Moussallam}. Clearly such a coupling plays no role
at tree level in the determination of the $L_i$'s whereas, in
principle, it will contribute at one loop. See also Ref.
\cite{Knecht-Nyffeler} for some other related limitations of this
resonance Lagrangian.

In our view the scale $\mu^*$ is reminiscent of, for instance, the
scale $M_X$ of gauge coupling unification in GUTs\cite{GUT}. In
fact, more physically meaningful than the value of $\mu^*$ itself
are equations such as, e.g., $L^{\mathcal{R}}_{i}(\mu^*)=0$ (for
all $i=1,...,10$), since they lead to relations among the
$L_{i}(\mu)$ at $\mu=\mu^*$ and therefore at all $\mu$. They can
be used as a guide in the search for a more predictive resonance
model. In this context one expects that even heavier resonances
than those in the Lagrangian (\ref{resonance}) are the ones which
give rise to the couplings $L^{\mathcal{R}}_{i}(\mu)$ upon
integration. Again we find in the framework of GUTs equations like
$\alpha_{\mathrm{SU_3}}(M_X)=\alpha_{\mathrm{SU_2}}(M_X)=\alpha_{\mathrm{U_1}}(M_X)$,
which are a good example of this type of relations.

A particularly interesting situation for its high predictive power
is what we could call the ``extreme'' version of resonance
saturation. This is when $L^{\mathcal{R}}_{i}(\mu)=0$ (for all
$i=1,...,10$) and for all $\mu$. In fact, just as $L_{10}(\mu)$
is obtained through a matching condition on the $\Pi_{LR}$
function which is an order parameter of spontaneous chiral
symmetry breaking, so are all the other $L_{i}(\mu)$ obtained
through corresponding matching conditions on certain QCD Green's
functions $\mathcal{G}$ which, because they are also order
parameters, vanish in the chiral limit to all orders. This
implies that all the $\mathcal{G}$'s have a finite and smooth
short-distance behavior. It is conceivable, and in our opinion
theoretically very appealing, that this finite ultraviolet
behavior be realized at the level of the resonance Lagrangian.
Restricted to the former Green's functions $\mathcal{G}$, the
resonance Lagrangian would then behave almost like renormalizable
and would predict, upon integration of the resonance fields, all
the $L_{i}(\mu)$ as a function of the resonance masses and
couplings. Clearly, we believe, this is the picture which gets
closest to the spirit of the work in Ref. \cite{Swiss1}. A
(surely oversimplified) sketch of the answer for $L_{10}(\mu)$ in
this picture could have been
\begin{equation}\label{sketch}
  L_{10}(\mu)= -\ \frac{1}{4}\left(\frac{15}{32 \pi^2 \sqrt{6}}\right)
  -\frac{1}{64\pi^2} \log\frac{\Lambda_{\chi}}{\mu}\ ,
\end{equation}
with $\Lambda_{\chi}$ a function of resonance masses and
parameters and $\Lambda_{\chi}\simeq 800$ MeV. We remark that the
coefficient in front of the logarithm should be the same as that
in Eq. (\ref{experiment}) \footnote{Notice how similar Eq.
(\ref{sketch}) is to the running of the electroweak angle in the
context of GUTs. For instance in $SU(5)$: $\sin^2\theta_{W}(\mu)=
\frac{3}{8}- \frac{55 \alpha}{24\pi}\log \frac{M_X}{\mu} $. In
this case the ``3/8'' is also a ratio of parameters in the
Lagrangian like our ``$-15/(32\pi^2\sqrt{6})\ $'' in Eq.
(\ref{prediction2}).}. However, we have shown that the resonance
interactions in the Lagrangian (\ref{resonance}) do not produce
this type of answer. The reason why our Eq. (\ref{result}) is
incompatible with the running of $L_{10}(\mu)$ in Eq.
(\ref{experiment}) and the condition
$L^{\mathcal{R}}_{10}(\mu)=0, \forall \mu$, is because the
resonance interactions in (\ref{resonance}) do not produce a
finite (i.e. $\mu$ independent) $\Pi_{LR}$ function. This is not
to be unexpected as (\ref{resonance}) lacks the right
short-distance properties it should
have\cite{Moussallam,Knecht-Nyffeler}. The dynamical challenge
clearly will be to incorporate all these short-distance
properties in a resonance Lagrangian which becomes more
ultraviolet convergent and yields finite answers for all the above
mentioned Green's functions $\mathcal{G}$; consequently predicting
all the $L_i(\mu)$ in this manner.

To conclude, we hope to have illustrated how one could use
large-$N_c$ in the context of a resonance Lagrangian to test in a
well-defined way the idea of resonance saturation at the quantum
level. Although our resonance Lagrangian  (\ref{resonance}) cannot
be considered fully realistic, it should be clear that a similar
analysis to the one presented here could be performed should a
more complete resonance Lagrangian of QCD be available. In this
sense the present analysis is complementary to that of Ref.
\cite{Knecht-Nyffeler} in the quest for a resonance Lagrangian
capable of pushing to higher energies the range of validity of the
description of QCD in terms of meson degrees of freedom.

\vskip 0.5in

 We thank G. Ecker, M. Knecht, J. Portoles and E. de
Rafael for discussions and reading the manuscript. This work is
supported by CICYT-AEN99-0766 and by TMR, EC-Contract No.
ERBFMRX-CT980169 (EURODA$\phi$NE).

\end{document}